\documentstyle[aps,twocolumn,prl,psfig]{revtex}
\begin{document}
\twocolumn[\hsize\textwidth\columnwidth\hsize\csname@twocolumnfalse\endcsname
\begin{title}
{\bf Spontaneous pattern formation in driven nonlinear lattices}
\end{title}

\author{Andrea Vanossi$^{1,2,4}$, K.~{\O}. Rasmussen$^1$, A.~R. Bishop$^1$,
  Boris A. Malomed$^3$, and V. Bortolani$^4$}
\address{$^1$Center for Nonlinear Studies and Theoretical Division,
Los Alamos National Laboratory,
Los Alamos, New Mexico 87545\\$^2$Dipartimento di Fisica, Universit\`a di
  Bologna, V.le Berti Pichat 6/2, I-40127, Bologna,
  Italy\\$^3$Department of Interdisciplinary Studies, Faculty of
  Engineering, Tel Aviv 69975, Israel\\$^4$INFM e Dipartimento di
  Fisica, Universit\`a di Modena, Via Campi 213/A, 41100 Modena, Italy}
\date{\today}
\maketitle
\begin{abstract}
  We demonstrate the spontaneous formation of spatial patterns in 
  a damped, ac-driven cubic Klein-Gordon
  lattice. These
  patterns are composed of arrays of intrinsic localized modes
  characteristic for nonlinear lattices. We analyze the modulation 
  instability leading to this spontaneous pattern formation. Our
  calculation of the modulational instability is applicable in one 
  and two-dimensional lattices, however in the analyses of the
  emerging patterns we concentrate particularly on the two-dimensional case.
\end{abstract}
\vspace*{10 mm}
]

Complex spatial patterns are often observed in systems driven away
from equilibrium \cite{Cross}. Typically, the patterns emerge when
relatively simple systems are driven into unstable states that will
deform dramatically in response to small perturbations. As the
patterns are arising from an instability, the pattern-forming behavior
is likely to be extremely sensitive to small changes in system
parameters. The description of deterministic pattern forming
systems is typically accomplished in the form of partial differential
equations such as the Navier-Stokes equations for fluids and
reaction-diffusion equations for chemical systems. These
phenomena have primarily been studied exclusively in continuum systems
such as hydrodynamical, optical, chemical systems and liquid crystals, although
more recently pattern formation of similar type have also been
reported in periodically vibrated granular media \cite{austin}.
  
Complementing the development of the theoretical understanding of 
pattern formation in continuum systems, the localized mode forming ability of
{\em discrete} lattices has also received significant recent
attention. There is now a fairly complete understanding of the existence
and stability of localized structures (often referred to as intrinsic
localized modes (ILM's) or discrete breathers) in a variety of nonlinear
lattices, undriven\cite{aubry} as well as driven\cite{jose}. It
is fair to claim that these collective patterns are well understood while
the process of their creation and interaction remains relatively unexplored.
  
  In the present communication we study the pattern forming
  abilities of a damped and periodically driven nonlinear lattice. 
  Specifically, we demonstrate how a driven nonlinear (cubic) 
  Klein-Gordon lattice forms a variety of patterns
  via modulational instabilities. 
  We analyse the modulational instabilities and show how these 
  relate to the length-scale of the patterns that are formed. Normally,
  the spatial extend (characteristic length-scale) of ILM's is directly 
related to the frequency of the ILM's \cite{aubry}. However, in our case 
the (generally different) length-scale emerging from the instability may lead to a 
length-scale competion, the results of which we will explore.
  
  {\em Model and stability of homogeneous solution.} First, we study
  the Klein-Gordon lattice
\begin{eqnarray}
\ddot x_{n}+\gamma \dot x_n +\omega_0^2 x_n=\Delta_n x_{n}
+\lambda x^3_n+\epsilon \cos\omega t,
\label{eqs}
\end{eqnarray}
where $\gamma$ is the damping parameter, $\omega_0$ the natural
frequency of the oscillators, $\lambda=\pm 1$ the nonlinearity parameter and
finally $\epsilon$ is the amplitude of the ac-drive at frequency
$\omega$. In one dimension the nearest neighbor coupling is
$\Delta_n x_n=x_{n+1}-2x_n+x_{n-1}$\cite{2Dana}. The amplitude $A_0$ (and
phase $\delta_0$) of the spatially homogeneous solution $x_n=A_0\cos(\omega
t+\delta_0)$ of Eq.(\ref{eqs}) can, within the rotating wave
approximation, be shown to satisfy
\begin{equation}
A_0^2
\left (\gamma^2\omega^2+(\omega^2-\omega_0^2+\frac{3}{4}
\lambda  A_0^2)^2 \right)=\epsilon^2.
\label{A0}
\end{equation}
The amplitude $A_0$ of the response to a driving amplitude $\epsilon$
is shown for the {\em soft } ($\lambda=1$) potential in Fig. 1. 
\begin{figure}[h]
\vspace*{250pt}
\hspace{-25pt}
\includegraphics{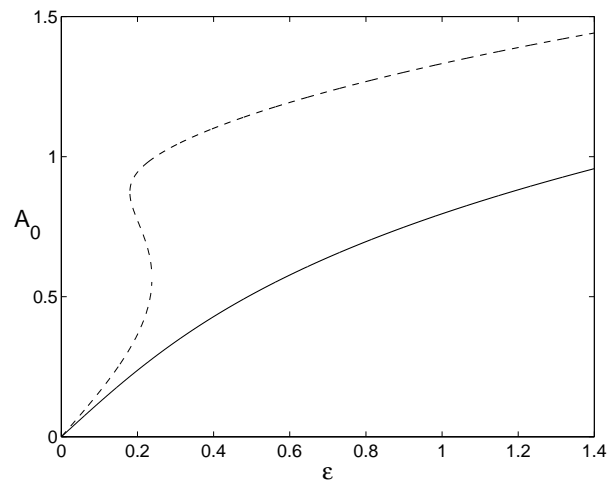}
\vspace*{-70pt}
\caption{Amplitude $A_0$ of response vs driving amplitude $\epsilon$ 
  of the ac-drive at frequency, $\omega=1.2\omega_0$ (solid line) and
  $\omega=0.8\omega_0$ (dashed line). Remaining parameters are
  $\omega_0=1.3$, $\gamma=0.15\omega_0$, and $\lambda=1$ (soft potential).}
\label{fig1}
\end{figure}
For $\omega <\omega_0$ (dashed line) three solutions are
possible, while for $\omega >\omega_0$ (solid line) a single solution
is possible. A similar picture is valid for the {\em hard}
($\lambda=-1$) potential except that the multiple solutions then appear in the
case $\omega >\omega_0$.

Analyzing the stability of the homogeneous solution with respect to
spatial perturbations, we introduce $x_n=y+z_n$ into Eq.(\ref{eqs}).
Assuming periodic boundary conditions, we may expand $z_n$ in its
Fourier components $z_n=\sum_k \exp( i k n) \xi_k(t)$, where the mode
amplitude $\xi_k(t)$ is then governed by
\begin{eqnarray}
\ddot \xi_k+\gamma\dot \xi_k+\omega_k^2\xi_k&=&\frac{3}{2}\lambda A_0^2 \left [
1+\cos(2\omega t +2 \delta_0) \right ] \xi_k
\label{HILL1} 
\end{eqnarray}
with $\omega_k^2=\omega_0^2+4\sin^2(k/2)$ denoting the linear dispersion
relation of the system.

Finally, the transformation $\xi_k(t)=\zeta_k(\omega
t+\delta_0)\exp(-\frac{\gamma}{2}(\omega t+\delta)) \equiv
\zeta_k(\tau)\exp(-\frac{\gamma}{2}\tau)$ reduces Eq.(\ref{HILL1}) to
a standard Mathieu equation
\begin{eqnarray}
\frac{d^2\zeta_k}{d\tau^2}+a\zeta_k-2q\cos(2\tau)\zeta_k=0,
\label{MATHIEU1}
\end{eqnarray}
where
\begin{equation}
a=\frac{1}{4\omega^2}\left ( 4\omega_k^2-6\lambda A_0^2-\gamma^2 \right )
\label{a}~~~~
\mbox{and}~~~
q=\frac{3\lambda A_0^2}{4\omega^2}.
\label{q}
\end{equation}
As is well-known \cite{ARN} the Mathieu equations exhibit
parametric resonances when $\sqrt{a} \simeq i$, where $i=1,2,3,...$.
The width of the resonance regions depends on the ratio $q/a$ (see,
e.g. \cite{stegun}). In the framework of Eq.(\ref{MATHIEU1}) the
extent of the primary resonance $a\simeq 1$ can easily \cite{stegun} be
estimated to be $(a-1)^2<q^2$. However, in the presence of the damping
$\gamma$ the resonance condition for Eq.(\ref{HILL1}) becomes
\begin{equation}
q^2> \frac{\gamma^2}{\omega^2}+(a-1)^2.
\label{basicreso}
\end{equation}
With $a$ and $q$ defined in Eqs.(\ref{a}), given
$\lambda,\gamma,\omega,\omega_0$, and $\epsilon$, this translates into
an instability band of certain wavenumbers $k$.

\begin{figure}[h]
\vspace*{250pt}
\hspace{-25pt}
\includegraphics{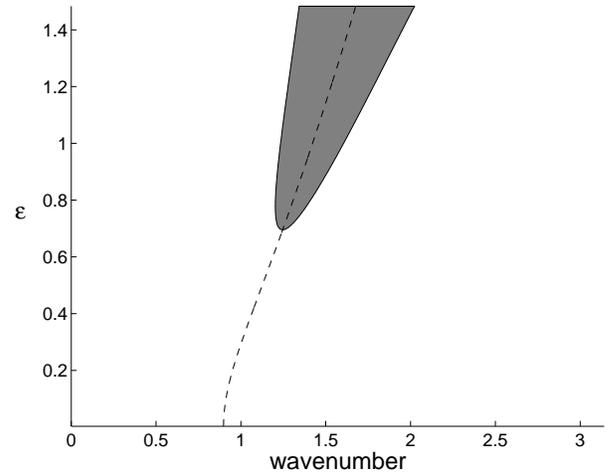}
\vspace*{-70pt}
\caption{Instability region (shaded) in the ($\epsilon$,$k$)-plane. Dashed line indicates the most unstable wavenumber
  $a=1$. Parameters corresponds to the solid line in Fig. 1.}
\label{fig2}
\end{figure}

Figure \ref{fig2} shows this instability band as given by
Eq.(\ref{basicreso}) for parameters corresponding to the solid curve
in Fig. \ref{fig1}. The shaded region is the band of wavenumbers that
are unstable according to Eq.(\ref{basicreso}) and the dashed line
indicates the most unstable wavenumber, {\em i.e. } $a=1$. The effect of the
damping clearly is to pinch off the instability region at a finite
driving $\epsilon >0$. Similar, using Eq.(\ref{basicreso}), instability 
regions can be determined for the solutions indicated by the dashed 
curve in Fig.\ref{fig1}.

We have verified the presence and location of the instability band by
direct simulations of Eq.(\ref{eqs}) starting from the initial
condition $x_n(t=0)=A_0\cos(\delta_0)+\eta_n,$ where, $\eta_n$,
represents a small ($|\eta_n| \ll A_0$) spatially random perturbation.
This initial condition injects energy into all wavenumbers and in the presence of 
an unstable region of wavenumbers the dynamics will enhance the 
energy content in this region and thereby identify the unstable region.

The above analysis is easily extended into the case of two spatial
dimensions\cite{2Dana}, the only required change being that the
dispersion relation now is
$\omega^2_{\vec{k}}=\omega_0^2+4\sin^2(k_x/2)+4\sin^2(k_y/2)$, where
the wavevector is $\vec{k}=(k_x,k_y)$. The instability in this case
appears on an annulus in the wavevector plane, with a radius given by
$a=1$ (see, Eq.(\ref{a})) and a width determined by
Eq.(\ref{basicreso}).

{\em Pattern formation.} Numerical simulations allow us not only to
verify the predicted instability band, but also to follow the full
nonlinear development and saturation initiated by the instability. In
particular, in regions of parameter space we obtain the spontaneous
formation of patterns of distinct spatial geometry. 
Although we have observed this phenomenon in one as well as in two
dimensions, in the present communication we focus on the 
two-dimensional system, where the pattern formation is very rich.
 
Although the dynamics show different features according to the specific
region of parameter space, it is possible to trace a typical 
behavior as 
follows: Initializing the system in the spatially homogeneous state 
described above with a small amount of randomness added, 
the instability sets in after a certain number of cycles of the
ac-drive, depending on the strength of the parametric resonance, 
{\em i.e. }on the value of $\sqrt{a} \simeq 1,2,3,...$.
Thereafter the system usually evolves through a sequence of 
different patterns (rhombi, stripes, etc.),
composed of localized regions of high amplitude oscillations,
before reaching a final configuration that may or may not result in a 
structure of definite symmetry.

Due to the sensitive response to very small
changes of the parameters, determining stability regions for the
different pattern geometries is a difficult task. 
However, in the case of a hard potential ($\lambda=-1$), Fig. 3 
shows a limited area of 
($\epsilon$,$\omega$)-space in which distinct spatial patterns emerge and
remain stable.

\begin{figure}[h]
\vspace*{250pt}
\hspace{-25pt}
\includegraphics{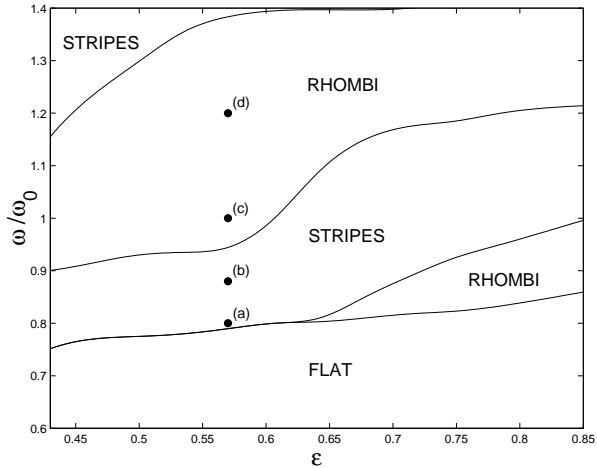}
\vspace*{-70pt}
\caption{Stability diagram showing the stability boundaries between the
  possible patterns in the ($\epsilon$,$\omega$)-plane. 
  Parameters are $\omega_0=0.75$, $\lambda=-1$ (hard
  potential), $\gamma=0.05\omega_0$. The four points refer to the
  specific spatial structures shown in Fig. 4.}
\label{fig3}
\end{figure}

This diagram is constructed following the full dynamics of the system
for thousands of cycles. As our study concentrates on patterns arising
from instabilities of the homogeneous solutions, we do not discuss
possible hysteretic behavior as is sometimes observed in
similar systems (see, e.g. Ref.[2]).

Figure 4 shows representative examples of 
the spontaneously emerging patterns corresponding to the points marked in Fig. 3.

\begin{figure}[h]
\vspace*{250pt}
\hspace{-25pt}
\includegraphics{fig4.ps}
\vspace*{-70pt}
\caption{(Color) Spatial patterns corresponding to the four points marked in
  Fig. 3 for $\epsilon=0.57$: (a) straight stripes ($\omega=0.8\omega_0$),
  (b) modulated stripes ($\omega=0.88\omega_0$), (c) rhombi ($\omega=1.0
  \omega_0$), (d) localized rhombi ($\omega=1.2\omega_0$). $x_{n,m}$
  is plotted along the vertical axes.}
\label{fig4}
\end{figure}

The patterns consist of localized regions of high amplitude 
oscillations (ILM's) residing on a background that oscillates 
at the frequency, $\omega$, of the ac-drive.
In all the considered cases we have observed the natural result that
patterns are only energetically sustained when the ILM,
$\omega_{ILM}$, and driving, $\omega$, frequencies are commensurate, 
{\em i.e. }$\omega_{ILM}=n\omega$, where $n$ is an integer. For the 
patterns displayed in Fig. 4, $n=2$.
Further, the motion of the ILM's is out of phase, {\em i.e. }, at 
the points in time where the background oscillation 
reaches its maximal excursion the ILM's obtain their minimal 
amplitude such that at these points the state is completely
homogeneous.

At a fixed driving $\epsilon$, for increasing values of the frequency $\omega$, as in Fig. 4, we can
observe the following behavior: Due to the presence of the damping 
$\gamma$, at sufficiently small $\omega$ the spatially homogeneous solution is 
stable towards all possible spatial modulations such that the {\em flat} state is sustained. However, for
values near point (a) the system becomes unstable with respect to certain 
spatial modulations and spatial
patterns in the form of large stripes emerge (Fig. 4(a)). Increasing the 
frequency, these stripes become thinner and denser and begin to show an increasingly 
clearer modulation (Fig. 4(b)). The characteristic length 
scale of these patterns is set by the size of the unstable $\vec{k}$-vector according to the above analysis.
The nonlinear character of the system results in a transition 
towards a more isotropic geometry (rhombic) as the driving frequency 
is increased further (Fig. 4(c)). As in the case of the stripes, stronger localization of the ILM's arranged in
the rhombic pattern (Fig. 4(d)) is observed for even larger driving frequencies. The angle between the 
sides of the rhombus unit cell varies but for the hard potential it is always 
close to $\pi/2$. For values of $\epsilon$ larger than
those displayed in Fig. 3, the final mesoscopic patterns of the system 
dynamics are spatially disordered much like the phenomena observed in
granular media \cite{austin}. 
It is important to realize that the 
length of the unstable wavevector determines the length-scales of the final patterns, while 
the symmetry of the patterns is determined by the nonlinear character of the system. 

As a result of the periodic boundary conditions, the length scale of
the emerging patterns must be commensurate with the system
size. However, we have observed that the 
existence of a band (variability in length and angle) 
of unstable $\vec k$-vectors (see Fig. 2 and related
discussion) allows continuous accommodation of this constraint except for the discontinuous  
changes in length scales occurring when it is energetically favorable for the system 
to add (or subtract) an additional stripe (or row of ILM's).

For the soft potential ($\lambda=1$) the variation of the amplitude $\epsilon$, and
the frequency, $\omega$ of the ac-drive is particularly problematic as 
the dynamics in this case can lead to the development of catastrophic instabilities as 
one or several oscillators overcome the finite barrier in the quartic 
potential.In all the cases we have been able to simulate, the early stage time
evolution of the system is characterized by the formation of ILM's 
regularly arranged in a square pattern. This spatial configuration,
which is sustained for up to hundreds of cycles, seems always to
suffer from a weak instability and eventually deforms into a rhombic
pattern, as shown in Fig. 5.
\begin{figure}[h]
\vspace*{250pt}
\hspace{-25pt}
\includegraphics{fig5.ps}
\vspace*{-70pt}
\caption{(Color) Snapshot of spontaneously formed pattern of ILM's. The red
  line represents a one-dimensional cut of the pattern array, 
  whose dynamics we have analyzed in detail (see Fig. 6). }
\label{fig5}
\end{figure}
Contrasting with the case of the hard potential, here in the soft
potential the 
angle between the sides of the rhombus unit cell is always close to
$2\pi/3$ (so almost hexagonal).

We now analyze this pattern more closely.  
The pattern shown in Fig. \ref{fig5} consists of ILM's spontaneously
organized into a regular rhombic pattern. The ILM's are spatially
localized and perform harmonic temporal oscillations (at the frequency,
$\omega$, of the ac-drive) and are therefore objects that are well
described \cite{aubry} in the literature. A particular feature of these
ILM's is that they reside on a background that oscillates at the same
frequency. To expose the dynamics of the ILM's more closely, we show in Fig.
\ref{fig6} a time sequence along a one-dimensional cut of the
two-dimensional system (the cut is indicated by the line in Fig.
\ref{fig5}).
It should be noted that in Fig. \ref{fig6} we have
removed the oscillations of the background in order to expose the ILM
dynamics most clearly.

Although we have focused on the ILM dynamics in the case of a soft potential, the features are 
analogous in the case of the hard potential and only the symmetry of the patterns is different.
\begin{figure}[h]
\vspace*{250pt}
\hspace{-25pt}
\includegraphics{fig6.ps}
\vspace*{-70pt}
\caption{ILM dynamics monitored along the direction indicated by the 
line in Fig. 5.}
\label{fig6}
\end{figure}

In summary we have studied the modulational instability in the damped
and ac-driven cubic nonlinear Klein-Gordon 
lattice. The analysis applies to one as well as two spatial dimensions. We have further demonstrated how these
instabilities lead to a variety of mesoscopic patterns of intrinsic
local modes. In the case of a hard potential we characterized
the patterns in a stability diagram and in the case of the soft potential we showed that the dynamics 
always result in a rhombic pattern with an angle close to $2\pi/3$. These rhombic patterns were 
never observed in the case of the hard potential. This difference in
the shapes of the rhombic patterns
in the hard and soft cases can be understood by exploiting the analogy between the changes in the steady states 
of a dissipative systems and phase transitions in systems at thermodynamic equilibrium (see. e.g. Ref. \cite{Malomed}).
In terms of this analogy the appearance of spatially periodic structures in a nonequilibrium system 
can be connected to a perturbation of the translational symmetry of thermodynamic states in equilibrium. A study based
on this philosophy was pursued in Ref.\cite{Malomed} with the results that the angle defining rhombic patterns 
is given by the coefficient of the cubic term in the system, which is 
precisely our observation. From the present analysis it appears that 
experimental 
studies of the pattern forming abilities of discrete systems present an excellent opportunity to study 
ILM's and their mutual interactions and mesoscopic patterning. For example, optical systems \cite{DH} and spin systems \cite{lai} 
appear good candidates for such studies.

Research at Los Alamos National Laboratory is performed under the auspice of 
the US DoE.


\begin{thebibliography}{10}
\bibitem{2Dana} The two-dimensional version of Eq.(\ref{eqs}) is
  $\ddot x_{n,m}+\gamma \dot x_{n,m} +\omega_0^2 x_{n,m}=
  x_{n+1,m}-4x_{n,m}+x_{n-1,m}+x_{n,m-1}+x_{n,m+1} +\lambda x^3_{n,m}
  +\epsilon \cos\omega t$, where the parameters have the
  interpretation given in connection with Eq. (\ref{eqs}).
\end{thebibliography}
\end{document}